\documentclass[%
reprint,
superscriptaddress,
amsmath,amssymb,
prl,
nobibnotes,
]{revtex4-2}

\usepackage[unicode=true,colorlinks=true,citecolor=blue,urlcolor=blue]{hyperref}

\usepackage{gensymb}
\usepackage{graphicx}
\usepackage{bm}

\usepackage{amsthm}
\usepackage{physics}
\usepackage{mathtools}

\allowdisplaybreaks[4]

\begin{document}

\title{Bulk-spatiotemporal vortex correspondence in gyromagnetic double-zero-index media}

\affiliation{%
Department of Physics, The Hong Kong University of Science and Technology, Hong Kong, China
}%
\affiliation{%
State Key Laboratory of Terahertz and Millimeter Waves and Department of Electrical Engineering, City University of Hong Kong, Hong Kong, China}
\affiliation{%
China State Key Laboratory of Radio Frequency Heterogeneous Integration, College of Electronics and Information Engineering, Shenzhen University, Shenzhen 518060, China
}%
\affiliation{%
Optoelectronics Research Centre, University of Southampton, Southampton SO17 1BJ, United Kingdom
}%

\author{Ruo-Yang Zhang
\textsuperscript{1}}
\thanks{These authors contributed equally.\label{equal contribution}}

\author{Xiaohan Cui
\textsuperscript{1,\ref{equal contribution}}}
\email[email: ]{xcuiad@connect.ust.hk}

\author{Yuan-Song Zeng\textsuperscript{2,\ref{equal contribution}}}

\author{Jin Chen\textsuperscript{2}}

\author{Wenzhe Liu\textsuperscript{1}}

\author{Mudi Wang\textsuperscript{1}}

\author{Dongyang Wang\textsuperscript{4}}

\author{Zhao-Qing Zhang\textsuperscript{1}}

\author{Neng Wang\textsuperscript{3}}%
\email[email: ]{nwang17@szu.edu.cn}

\author{Geng-Bo Wu\textsuperscript{2}}
\email[email: ]{bogwu2@cityu.edu.hk}

\author{C. T. Chan\textsuperscript{1}}%
\email[email: ]{phchan@ust.hk}

\begin{abstract}
Photonic double-zero-index media, distinguished by concurrently zero-valued permittivity and permeability, exhibit extraordinary properties not found in nature. Remarkably, the notion of zero-index can be substantially expanded by generalizing the constitutive parameters from null scalars to nonreciprocal tensors with nonzero matrix elements but zero determinants. Here, we experimentally realize such a new class of gyromagnetic double-zero-index metamaterials possessing both double-zero-index features and nonreciprocal hallmarks. As an intrinsic property, this metamaterial always emerges at a spin-1/2 Dirac point of a topological phase transition. We discover and rigorously prove that a spatiotemporal reflection vortex singularity is always anchored to the metamaterial’s Dirac point, with the vortex charge being determined by the topological invariant leap across the phase transition. This establishes a unique bulk-spatiotemporal vortex correspondence that extends the protected boundary effects into the time domain and exclusively characterizes topological phase transition points, setting it apart from any pre-existing bulk-boundary correspondence. Based on this correspondence, we propose and experimentally demonstrate a mechanism to deterministically generate optical spatiotemporal vortex pulses with firmly fixed central frequency and momentum, hence showing unparalleled robustness. Our findings uncover deep connections between zero-refractive-index photonics, topological photonics, and singular optics, opening the avenue for the manipulation of space-time topological light fields via the inherent topology of extreme-parameter metamaterials.

\end{abstract}

\maketitle
\newpage


Photonic zero-refractive-index (ZRI) media~\cite{ziolkowski2004Propagation,silveirinha2006Tunneling,liberal2017Nearzero,kinsey2019Nearzeroindex} are an exemplary family of extreme parameter metamaterials, which exhibit extraordinary optical properties and have wideranging applications in wave manipulation~\cite{liberal2017Photonic,ciattoni2017Efficient,liu2023Broadband} and nonlinear optics~\cite{suchowski2013Phase,alam2016Large}. 
In this family of metamaterials, a special member is the media with both zero permittivity ($\varepsilon$) and permeability ($\mu$), known as double-zero-index metamaterials (DZIMs)~\cite{huang2011Dirac,moitra2013Realization,li2015Onchip,cui2019Realization,xu2021ThreeDimensional,li2021Diraclike}. DZIMs are unique in that
they not only present a ZRI but also show a conical band intersection at the double-zero-index (DZI) frequency. This property unlocks a myriad of exclusive optical characteristics and functionalities, such as universal impedance matching and perfect transmission~\cite{huang2011Dirac,fang2016Klein,luo2015Unusual}, 
ZRI bound states in the continuum~\cite{minkov2018ZeroIndex,dong2021Ultralowloss},
and large-area single-mode lasing~\cite{chua2014Largerareaa,contractor2022Scalable}.

While the conventional DZIMs have null scalar $\varepsilon$ and $\mu$, it has been shown that by incorporating  anisotropic and nonreciprocal constitutive tensors with nonzero matrix elements but zero determinants, the realm of ZRI metamaterials can be substantially expanded~\cite{davoyan2013Theory,horsley2021Zerorefractiveindex,davoyan2013Optical,yang2023Magnetically}. In this work, we experimentally realize, for the first time, such a generalized DZIM, called gyromagnetic double-zero-index metamaterials (GDZIMs)~\cite{zhou2018Realization,wang2020Effective,feng2022Magnetooptical}, possessing a null scalar permittivity $\varepsilon_z$ and a nonzero gyromagnetic permeability tensor with a determinant of zero $\mathrm{det}(\tensor{\mu}) = 0$.
We theoretically and experimentally reveal that the GDZIMs not only maintain the prominent features of conventional DZIMs, but also possess exotic nonreciprocal characteristics that allow us to observe unprecedented wave phenomena in ZRI photonics.

\begin{figure*}[t!]
\includegraphics[width=0.66\textwidth]{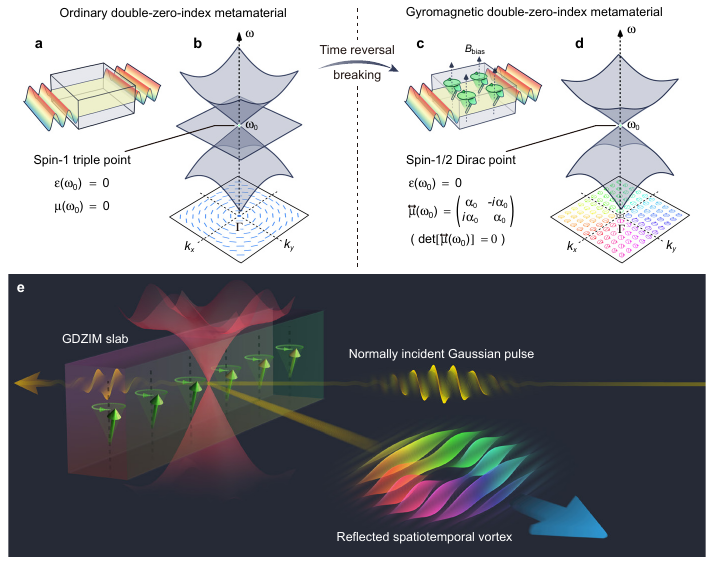}
\caption{\label{Fig-schematic} 
\textbf{Comparing the characteristics of gyromagnetic and ordinary double-zero-index materials.}
\textbf{a},\textbf{c}, Schematics of plane waves in homogeneous ordinary (left) and gyromagnetic (right) DZIMs.
The green arrows in (\textbf{c}) indicate the gyromagnetic precession of the magnetic moments around a DC bias magnetic field inside the GZIM. \textbf{b},\textbf{d}, The band structures near the double-zero-index frequency ($\omega_0$) form  (\textbf{b}) a spin-1 cone and (\textbf{d}) a spin-1/2 Dirac cone in ordinary and gyromagentic DZIMs, respectively. The bars and ellipses on the bottom planes represent the eigen polarization of the magnetic field $\mathbf{H}(\mathbf{k})$ of the upper band in the two kinds of DZIMs. In (\textbf{b}) an ordinary DZIM, the linearly polarized magnetic fields approach a V-point type singularity at the $\Gamma$ point ($\mathbf{k}=\mathbf{0}$) of the momentum space, while in (\textbf{c}) a GDZIM, the eigen fields approach circular polarization $|C\rangle=(\hat{\mathbf{x}}-i\hat{\mathbf{y}})/\sqrt{2}$ (a C point) at $\Gamma$. The colors of polarization ellipses in (D) represent the phase of the eigenfield projected to the circular basis: $\phi=\arg\langle C|\bf{H}(\bf{k})\rangle$. \textbf{e}, Illustration of a 2D Gaussian pulse normally impinging upon a GDZIM slab at the Dirac cone frequency and the normal reflection of the pulse creating a spatiotemporal vortex. 
}\vspace{-6pt}
\end{figure*}

The resemblance and distinction between ordinary DZIMs (Fig.~\ref{Fig-schematic}a,b) and GDZIMs (Fig.~\ref{Fig-schematic}c,d) are first encoded in their band structures near the DZI frequency $\omega_0$. 
In a two-dimensional (2D) ordinary DZIM with linear material dispersion near $\omega_0$,
i.e., $\varepsilon(\omega) =c_\varepsilon(\omega-\omega_0)$ and
$\mu(\omega) =c_\mu(\omega-\omega_0)$,
the frequency bands exhibit a so-called ``Dirac-like'' triple crossing between a cone and an irremovable flat band at the $\Gamma$ Point ($\mathbf{k}=0$) of momentum space (Fig.~\ref{Fig-schematic}b). Likewise, a 2D GDZIM for transverse magnetic (TM) waves can always be characterized by linearly dispersive constitutive parameters around $\omega_0$:
\begin{equation}\label{GDZIM constitutive parameters}
\begin{aligned}
\varepsilon_z(\omega) &=c_\varepsilon(\omega-\omega_0),\\
\tensor{\mu}_{xy}(\omega) &=\begin{pmatrix*}[r]
\alpha_0 & -i\alpha_0 \\ i\alpha_0 & \alpha_0
\end{pmatrix*}+\begin{pmatrix*}[r]
c_{\mu_d} & -ic_{\alpha} \\ ic_{\alpha} & c_{\mu_d}
\end{pmatrix*}(\omega-\omega_0),
\end{aligned}
\end{equation}
which gives rise to a spin-1/2 Dirac cone dispersion pinned at $(\omega,\mathbf{k})=(\omega_0,\mathbf{0})$ (Fig.~\ref{Fig-schematic}d).
Notwithstanding the analogous conical shape, the Dirac-like cone in ordinary DZIMs 
has a spin-1 nature and is thus fundamentally different from a true Dirac cone of spin-1/2~\cite{fang2016Klein}. Indeed, in bosonic systems, time-reversal symmetry inherently prohibits a 2D spin-1/2 Dirac point from existing at time-reversal-invariant momenta (see Supplementary Note S1.B). Therefore, the appearance of the centered spin-1/2 Dirac point in GDZIMs directly signifies time-reversal breaking and implies nontrivial nonreciprocal effects~\cite{wang2020Effective,feng2022Magnetooptical}. For instance, the eigen magnetic fields in GDZIMs converge to an in-plane circular polarization at $\Gamma$, as a result of the gyromagnetic precession of the magnetic moments induced by the external DC bias magnetic field. This highlights the specific preference of GDZIMs for a certain out-of-plane polarization of the transverse optical spin $\mathbf{S}_\mathrm{opt}=\frac{\mu_0}{4\omega_0}\mathrm{Im}(\mathbf{H}^*\times\mathbf{H})$ 
for waves propagating in any in-plane direction.

\begin{figure*}[t!]
\includegraphics[width=\textwidth]{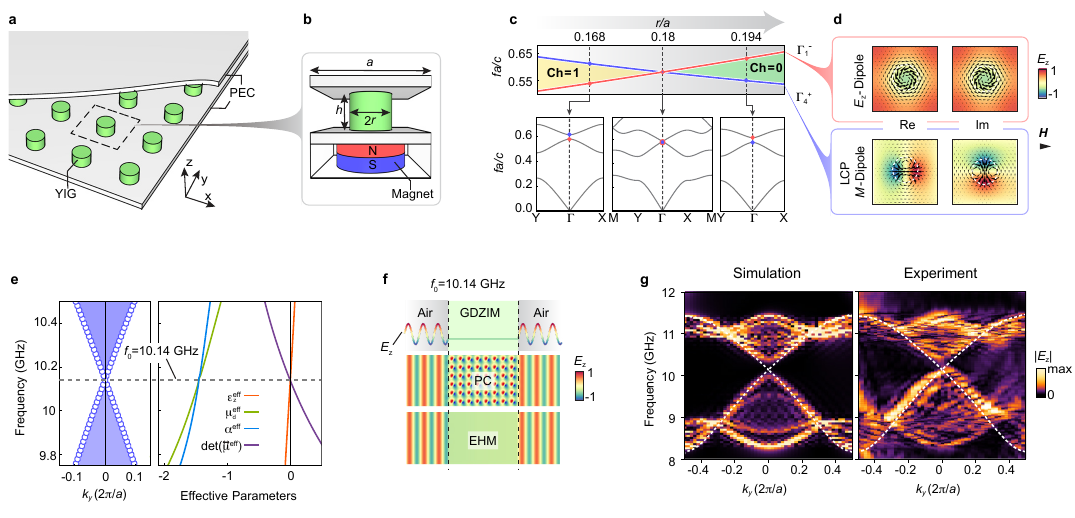}
\caption{\label{Fig-PC} 
\textbf{Experimental realization of GDZIM.}
\textbf{a}, Schematic of the gyromagentic PC. \textbf{b}, Unit cell of the PC with lattice constant $a=17.2\,\mathrm{mm}$; a YIG rod (height: $h=4\,\mathrm{mm}$, radius: $r=3.1\,\mathrm{mm}$) is sandwiched between two PEC parallel plates; a permanent magnet under the YIG rod creates an effective bias magnetic field $H_0=900\,\mathrm{Oe}$ in the $z$ direction. \textbf{c}, Upper panel: Band inversion of the two $\Gamma$-point modes on the second and third TM bands of the PC controlled by the filling ratio ($r/a$) of YIG rods ($\varepsilon_z = 13$, $\mu_d=0.865$, $\alpha=-0.54$) . Lower panel: Band structures along high-symmetry lines for three PCs in the trivial phase, at phase transition point, and in the Chern insulator phase, respectively. \textbf{d}, Profiles of the two $\Gamma$-point modes ($\Gamma^-_1$: out-of-plane electric dipole and $\Gamma^+_4$: in-plane left circularly polarized (LCP) magnetic dipole). \textbf{e}, Retrieved effective parameters and band dispersion (blue circles) of the effective medium near the Dirac frequency $f_0=10.14\,\mathrm{GHz}$. \textbf{f}, Simulation of plane wave propagation through the PC and effective medium slabs at  $f_0$. \textbf{g}, Simulated and measured projected bands of a PC with $30 \times 10$ unit cells.  Red dashed lines represent the anticipated bulk Dirac cone along $k_y=0$. 
}\vspace{-5pt}
\end{figure*}

\begin{figure*}[t!]
\includegraphics[width=0.66\textwidth]{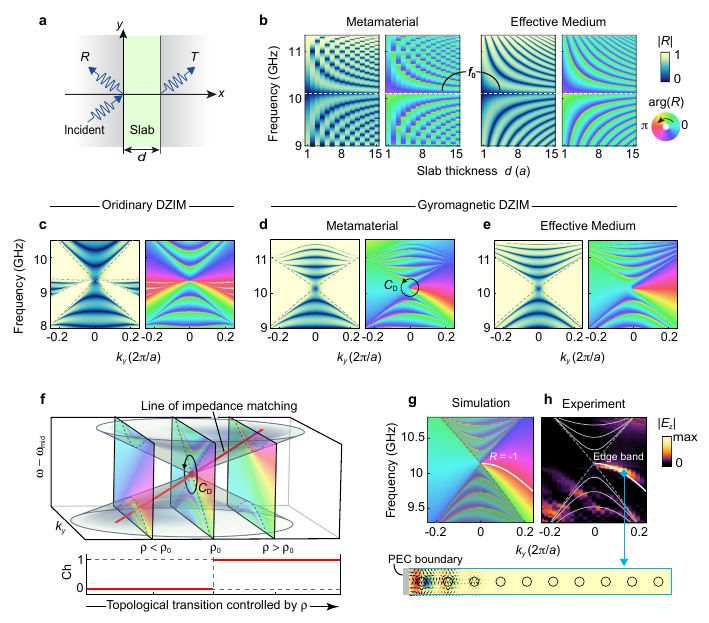}
\caption{\label{Fig-3} 
\textbf{Ultrarobust reflection phase vortex and its topological origin.}
\textbf{a}, Diagram of plane wave reflection by a GDZIM slab. \textbf{b}, Normal incidence reflection spectra by GDZIM slabs made of (left) metamaterials and of (right) effective homogeneous media (EHM), changing with slab thickness $d$. White dashed line: Dirac frequency $f_0=10.14\,\mathrm{GHz}$. \textbf{c}, Reflection amplitude (left) and phase (right) of an ordinary DZIM slab with a spin-1 Dirac-like cone. Parameters of DZIM: $\varepsilon_z=13$, $\mu = 1$, $r_c =3.4 \mathrm{mm}$ and $a =17.2 \mathrm{mm}$.  \textbf{d},\textbf{e}, Reflection spectra of (\textbf{d}) a GDZIM slab and of (\textbf{e}) the corresponding EHM slab. The slab thickness of (\textbf{c}-\textbf{e}): $d=6a$. \textbf{f}, Evolution of the reflection spectra during the topological phase transition process of a gyromagnetic near-DZIM slab, which happens to be an exact GDZIM at the critical gap closing point $(\rho,k_y,\omega)=(\rho_0,0,\omega_0)$. White cone: edges of projected bulk bands. Red line: loci satisfying impedance matching and $R=0$. \textbf{g}, Simulated reflection phase of a 10-layer metamaterial slab, where the equi-phase contour $\arg(R)=\pi$, i.e., $R=-1$, predicts the band of edges states localized on a PEC boundary. \textbf{h}, Experimentally measured edge band compared with theoretical prediction (white line). Bottom inset: eigen electric (colormap) and magnetic  (black vectors) fields of the edge state at $k_y=0.138$ in the supercell.
}\vspace{-5pt}
\end{figure*}

GDZIMs inherit the salient characterisitc of DZIMs, 
namely the ability to match impedance with any surrounding medium, and hence support perfect transmission (zero reflection) at the DZI frequency $\omega_0$ for normally incident plane waves. However, unlike the bilaterally symmetric reflection by ordinary DZIMs, the nonreciprocity of a GDZIM slab allows for an asymmetric reflection about the normal axis ($k_\parallel=0$), resulting in a spatiotemporal reflection phase vortex emerging and firmly anchored at the projection of the Dirac point, $(\omega,k_\parallel)=(\omega_0,0)$. 
Particularly, we uncover that a GDZIM always resides at the topological transition point of photonic Chern insulator phases, and the quantized charge of the reflection vortex is precisely determined by the variation of bulk Chern number across the phase transition, hence establishing a novel bulk-boundary correspondence (BBC).
This vortical reflection spectrum further ensures that when a Gaussian pulse with a central frequency of $\omega_0$ normally impinges upon the slab, the reflected wave always forms a spatiotemporal vortex pulse (STVP) (Fig.~\ref{Fig-schematic}e). Optical spatiotemporal vortices~\cite{jhajj2016Spatiotemporal,bliokh2021Spatiotemporal,hancock2019Freespace,chong2020Generation,liu2024Spatiotemporal,gui2021Secondharmonic,hancock2021Secondharmonic,Wang:21,zhang2023Topologically,liu2024Exploiting,che2024Generation,huo2024Observation,ni2024ThreeDimensional}, which carry transverse orbital angular momenta (OAM), have attracted considerable attention due to their potential applications in optical tweezers, superresolution imaging, and optical information processing. 
Compared to other methods of synthesizing STVPs, such as using customized spatial light modulators~\cite{hancock2019Freespace,chong2020Generation,liu2024Spatiotemporal} or resonance-based metasurfaces~\cite{Wang:21,zhang2023Topologically,liu2024Exploiting,che2024Generation,huo2024Observation,ni2024ThreeDimensional}, the current approach stems from the intrinsic Dirac cone topology of a homogenous GDZIM and possesses exceptional robustness in that the central frequency and wavevector of the generated STVPs are independent of the thickness of the GDZIM slab, the refractive index of the background medium, and even the crystal cutting directions of the realistic metamaterials.

We will now show how to design and realize the GDZIMs, as well as the theory and experiments of ultrarobust generation of STVPs in the microwave regime.

\vspace{-15pt}
\subsection{Design and realization of GDZIMs}\vspace{-10pt}
To realize the GDZIMs, we devise a 2D gyromagnetic photonic crystal (PC) (Figs.~\ref{Fig-PC}a,b) comprised of a square lattice of Yttrium iron garnet (YIG) rods sandwiched within a metallic parallel-plate waveguide. 
In each unit cell, a permanent magnet is placed right below the lower metallic plate, which produces a vertical DC bias magnetic field $B_\mathrm{bias}$ and magnetizes the YIG pillar above. 
Here, we focus on the second and third TM bands of the PC in Fig.~\ref{Fig-PC}c, on which the two $\Gamma$-point modes follow the irreducible co-representations $\Gamma^-_1\oplus\Gamma^+_4$ 
of the magnetic little co-group $4/mm'm'$. As shown in Fig.~\ref{Fig-PC}d, these two modes correspond to an out-of-plane electric dipole and an in-plane circularly polarized magnetic dipole, respectively, and therefore conform with the two eigen-polarizations ($\mathbf{E}\propto \hat{\mathbf{z}}$ and $\mathbf{H}\propto(\hat{\mathbf{x}}-i\hat{\mathbf{y}})/\sqrt{2}$) in a homogeneous GDZIM. The gap between the two bands can be continuously tuned by the filling ratio ($r/a$) of the PC.  
In our experimental setup, we fix $B_\mathrm{bias}$ and the size of the YIG pillars, and adjust the lattice constant $a$ to achieve the critical point of gap closing. Then, an accidentally degenerate spin-1/2 Dirac point~\cite{liu2020Observation} emerges at the center of the Brillouin zone (Fig.~\ref{Fig-PC}c), signifying the topological transition between a photonic Chern insulator phase~\cite{haldane2008Possible,wang2009Observation} with a nontrivial gap Chern number $\mathrm{Ch}=1$ and a trivial phase with $\mathrm{Ch}=0$. In this way, both the Dirac cone dispersion and the mode symmetry suggest that the PC behaves as a GDZIM at the topological phase transition point.

Employing the boundary effective medium (BEM) approach~\cite{wang2020Effective} (see Supplementary Note S1.C), we can rigorously retrieve the homogenized effective constitutive parameters of the PC near the Dirac point: $\varepsilon_z^\mathrm{eff}(\omega)$ and $\tensor{\mu}^\mathrm{eff}(\omega)=\begin{pmatrix*}[r]
    \mu_d^\mathrm{eff} & -i\alpha^\mathrm{eff} \\ i\alpha^\mathrm{eff} & \mu_d^\mathrm{eff}
\end{pmatrix*}$, as illustrated in Fig.~\ref{Fig-PC}e. At the Dirac frequency $f_0=\omega_0/(2\pi)=10.14$ GHz, we observe that both the 
effective permittivity and the determinant of effective permeability tensor reduce to zero: 
$\varepsilon_z^\text{eff}(\omega_0)=0$ and $\mathrm{det}(\tensor{\mu}^\text{eff}(\omega_0) )=0$, whereas the diagonal and gyromagnetic components of $\tensor{\mu}^\text{eff}$ remain finite and exhibit an identical value 
$\alpha^\text{eff}(\omega_0)=\mu_d^\text{eff}(\omega_0)=\alpha_0=-1.47$, which precisely reproduce the GDZIM properties described by Eq.~\eqref{GDZIM constitutive parameters}. 
In Fig.~\ref{Fig-PC}f, we observe no spatial phase change in the field within the GDZIM region.
The remarkable resemblance of field distributions for the PC and effective medium validates the efficacy of the BEM methodology. 

We have conducted experimental detection for the Dirac cone band dispersion of the PC. 
Figure~\ref{Fig-PC}g presents the measured projected band structure alongside the full-wave simulated results obtained using an identical configuration to the experiment (see Supplementary Note S1.A). Their good agreement substantiates that the constructed PC indeed achieves a GDZIM at the anticipated frequency $f_0=10.14\,\mathrm{GHz}$.

\vspace{-15pt}
\subsection{Ultrarobust reflection phase vortex}\vspace{-15pt}
The unique constitutive parameters and band structure of GDZIMs endow them with exceptional transmission and reflection peculiarities. To see this, we consider a TM plane wave incident upon a GDZIM slab with thickness $d$ embedded in a background medium with isotropic parameters $\varepsilon_1$ and $\mu_1$, as illustrated in Fig. \ref{Fig-3}a.
For normal incidence $(k_y=0)$, the reflection coefficient satisfies (see Supplementary Note S2)
\begin{equation}\label{Rnormal}
    R\propto \qty(\frac{e^{2i k^\mathrm{eff}_{x} d}-1}{k_x^\mathrm{eff}}) \qty(\varepsilon_1\tilde\mu^\mathrm{eff}-\varepsilon_z^\mathrm{eff}\mu_1) ,
\end{equation}
with $\tilde{\mu}^\mathrm{eff}=\det(\tensor{\mu}^\mathrm{eff})/\mu^\mathrm{eff}_d$ and $k_x^\mathrm{eff}$ denoting, respectively, the virtual scalar permeability and the $x$-component of wavevector in the GDZIM. 
Equation~\eqref{Rnormal} shows that perfect transmission ($R=0$) through the slab occurs, if either of the two terms on the right equals zero. The first term reduces to zero at 
Fabry-Perot (FP) resonances which depend on the slab thickness. 
In contrast, the vanishing of the second term defines the generalized impedance matching condition for gyromagnetic materials: 
$
\varepsilon_1\tilde\mu^\mathrm{eff}=\varepsilon_z^\mathrm{eff}\mu_1
$.
In particular, at the DZI frequency $\varepsilon_z^\mathrm{eff}(\omega_0)=\tilde{\mu}^\mathrm{eff}(\omega_0)=0$, the impedance matching is universally established for arbitrary values of $\varepsilon_1$, $\mu_1$. Therefore, 
the GDZIMs inherit the capability as ordinary DZIMs to match impedance with a medium of any parameters and arbitrary thickness~\cite{huang2011Dirac}.

Figure~\ref{Fig-3}b presents the normal incidence reflection spectra for slabs of various thicknesses made from the metamaterials shown in Fig.~\ref{Fig-PC}a and compared with the spectra of the effective homogeneous media retrieved. 
In both cases, a reflection zero  appears consistently at the Dirac frequency (white dashed lines) irrespective of the slab thickness, validating the unconditional impedance matching feature of GDZIMs. 
And all other reflection zeros are induced by the FP effect, causing their frequencies to change with the thickness. 
A more crucial difference between the two types of reflection zeros is that only the DZIM-induced reflection zeros have the potential to generate reflection phase vortices, because only this type of zeros are isolated singular points when considering oblique incidence, whereas FP-induced reflection zeros always join into curves along $k_y$, as evidenced by Figs.~\ref{Fig-3}c-e.

Although GDZIMs and ordinary DZIMs both support universal impedance matching for normal incidence, the off-normal reflection feature of GDZIMs differs significantly from that of conventional DZIMs by virtue of nonreciprocity. 
The presence of reciprocity and mirror-$y$ symmetry of an ordinary DZIM slab imposes two constraints on the scattering matrix (see Supplementary Note S3): $S(k_y,\omega)^T=S(-k_y,\omega)$ and $S(k_y,\omega)=S(-k_y,\omega)$, both of which protect
the reflections bilaterally symmetric about the $k_y=0$ axis: $R(-k_y,\omega)=R(k_y,\omega)$, as exemplified in Fig.~\ref{Fig-3}c. Consequently, the winding of reflection phase around the reflection singularity (projection of Dirac-like point) must be zero, rendering the singularity topologically fragile and susceptible to elimination by imperfections. However, for GDZIM slabs, both reciprocity and mirror-$y$ symmetry are broken due to the DC bias magnetic field. Thus, as depicted by the reflection spectra of both metamaterials (Fig.~\ref{Fig-3}d) and effective media (Fig.~\ref{Fig-3}e), the reflections from a GDZIM slab are generically asymmetric about the normal axis, which leads to    
a deterministic formation of a topological reflection phase vortex encircling the isolated reflection singularity anchored at the projected Dirac point,  $(k_y,\omega)=(\omega_0,0)$, in the $(k_y,\omega)$-plane.

The GDZIM-induced reflection phase vortex is topologically stable as protected by quantized phase winding number around it, $\nu(R,C_D)=\frac{1}{2\pi}\oint_{C_D} d\arg(R)=\pm1$~\cite{ni2021Multidimensional}. 
But beyond that, it exhibits unparalleled robustness, surpassing the usual spectral phase singularities which can merely persist but generally move with varying parameters. In contrast, the frequency and wavevector of the GDZIM-induced vortex center are solidly pinned, regardless of the background medium, the slab thickness, or even the specific crystal plane along which the metamaterials are truncated (see examples in Supplementary Notes S2.C,D).

\begin{figure*}[t!]
\includegraphics[width=0.66\textwidth]{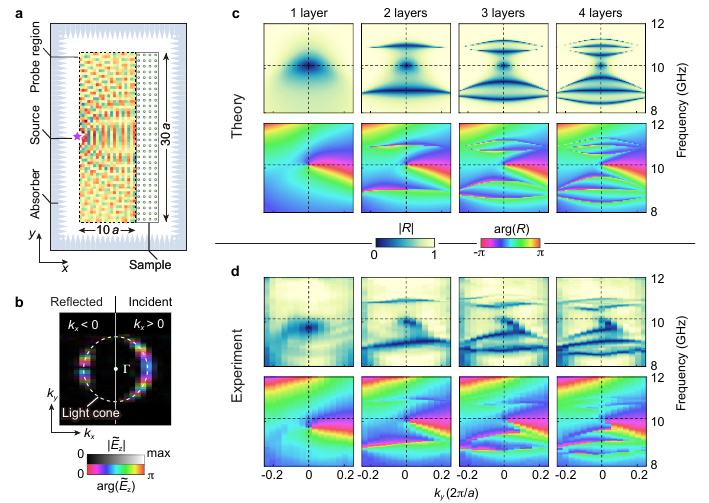}
\caption{\label{Fig-4} 
\textbf{Experimental measurement of ultrarobust reflection phase vortices.}
\textbf{a}, Diagram of reflection phase measurement setup. \textbf{b}, The Fourier spectra of the incident $E^\text{in}_z = \tilde{E}_z (k_x>0)$ and reflected field $E^\text{r}_z = \tilde{E}_z (k_x<0)$ measured at $10$ GHz. \textbf{c},\textbf{d} The theoretical and experimental results for the phase and amplitude of the reflection spectrum $R(k_y,\omega)$ for GDZIM with different layers. The reflection vortices are fixed at the Dirac frequency $f_0=10.14$GHz (black dashed line) for different layers. 
}\vspace{-5pt}
\end{figure*}

\vspace{-15pt}
\subsection{Bulk-spatiotemporal vortex correspondence}
\vspace{-10pt}
The remarkable stability of the GDZIM-induced reflection vortex can be further traced to a more profound bulk topological origin, i.e., a new type of BBC linking the topological phase transition at the accidental Dirac point and the emergence of the spatiotemporal reflection phase vortex of a 2D homogeneous medium with finite thickness. Figure~\ref{Fig-3}f sketches the evolution of the reflection spectrum of a metamaterial slab that undergoes a topological transition, modulated by a variable structural parameter $\rho$ of the metamaterial. As $\rho$ varies, the reflection vortices coalesce into a singularity line in the $(\rho,k_y,\omega)$ space. On the present occasion, this line coincides with the impedance matching frequencies, denoted by $\omega_\mathrm{IM}$, obeying $\varepsilon_1\tilde{\mu}^\mathrm{eff}(\omega_\mathrm{IM},\rho)=\varepsilon_z^\mathrm{eff}\mu_1(\omega_\mathrm{IM},\rho)$.
In particular, we have rigrously proved in Supplementary Note S4 that the discontinuous jump in Chern number across the phase transition enforces the line of reflection singularities to pass through the critical gap closing point $(\rho_0,0,\omega_0)$. Consequently, a spatiotemporal reflection phase vortex arises almost invariably at the surface projection of the accidental Dirac point, and the topological charge carried by the vortex is definitely determined by the change of bulk Chern number across the topological phase transition, together with the slope of the singularity line:
\begin{equation}\label{BBC}
    \nu(R,C_D)=\mathrm{sgn}\qty[\left.\frac{d \Delta\omega_\mathrm{IM}}{d\rho}\right|_{\rho_0}]\qty(\mathrm{Ch}_{\rho>\rho_0}-\mathrm{Ch}_{\rho<\rho_0}),
\end{equation}
where  
$\Delta\omega_\mathrm{IM}=\omega_\mathrm{IM}-\omega_\mathrm{mid}$ with $\omega_\mathrm{mid}(\rho)$ denoting the mid-gap frequency as a function of $\rho$.

Compared with any known BBCs~\cite{ozawa2019Topological,hu2015Measurement,liu2020Quantized,cheng2020Vortical}, the new correspondence bears three fundamental distinctions. First, the bulk topology is not attributed to a single phase, rather is decided by the topological transition between two phases. Second, unlike the usual scenarios where the boundary effects are well-defined only for a semiinfinite or large enough bulk, the current boundary reflection inherently requires that the thickness of the continuous bulk medium is finite and can even take on very small values. 
Third, as elucidated later, the new BBC intrinsically leads to a spatiotemporal boundary effect, i.e., the ultrarobust generation of STVPs,  setting it apart from any stationary topological boundary effects predicted by conventional BBCs.

As an application of this BBC in the sufficiently thick limit, the spatiotemporal reflection phase vortex elucidates why a bulk Dirac point nearly always induces a band of edge states that are localized on a hard-wall boundary and connect to its projection point. Specifically, when an insulated boundary encases the left end of a GDZIM, with a free-space reflection coefficient $R_\mathrm{bdy}(k_y,\omega)=e^{i\phi_\mathrm{bdy}}$, the phase matching condition for edge states subsisting at this boundary is $R_\mathrm{bdy}(k_y,\omega)R(k_y,\omega)=1$~\cite{cui2022Photonic}. Therefore, as long as the boundary is time-reversal invariant, say, $\phi_\mathrm{bdy}(k_y,\omega)$ takes any constant value, the vortical reflection phase of the GDZIM ensures that the phase matching can always be satisfied along a half-curve terminating at the vortex center, i.e., the projected Dirac point. Figure~\ref{Fig-3}g illustrates the examples of a PEC boundary. Since $R_\mathrm{bdy}=-1$ for PEC, 
the edge states on the boundary should disperse along the constant reflection phase curve $\arg(R)=\pi$, 
which precisely predict the actual edge band obtained through simulations. The experimental test of the edge states on a PEC boundary  (Fig.~\ref{Fig-3}h) further verifies the approach.

\begin{figure*}[t!]
\includegraphics[width=\textwidth]{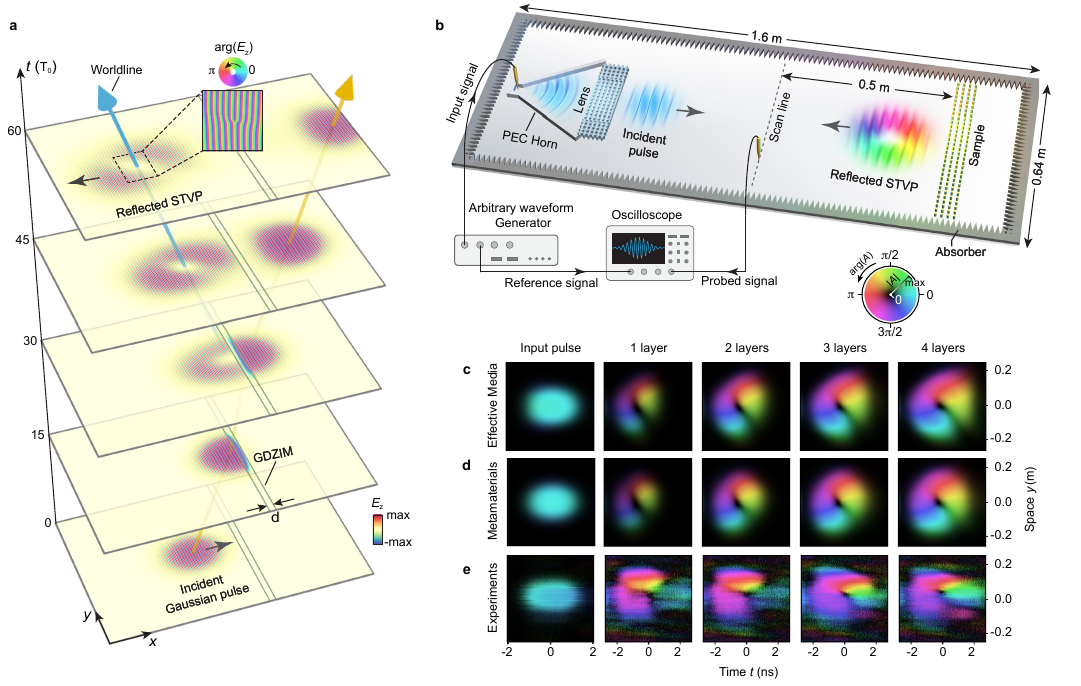}
\caption{\label{Fig-5} 
\textbf{Generation and observation of spatiotemporal vortex pulses.}
\textbf{a}, Time slices showing the generation of a STVP by reflecting a normally incident Guassian pulse upon a homogeneous GDZIM slab ($d=4a$). The blue and yellow arrows represent the centroid wordlines of the reflected STVP and of the transmitted and incident pulses, respectively. $T_0=2\pi/\omega_0$ denotes the temporal period corresponding to the DZI frequency $\omega_0$. \textbf{b}, Schematic of the experimental setup for generating and detecting STVPs. The scan line (gray dashed) is $0.5$m away from the PC samples (Size: $N\times 30$ units, $N=1$ to 4 layers). \textbf{c}-\textbf{e}, Envelopes and phases ($A(y,t)=E_z(y,t)e^{i\omega t}$) on the scan line of input pulse and reflected pulses corresponding to samples of 1 to 4 layers, which are obtained by (\textbf{c}) theoretical calculation of effective homogeneous medium slabs (thickness: $d=N\,a$), by (\textbf{d}) numerical simulations of metamaterial slabs, and by (\textbf{e}) experimental measurements. 
}\vspace{-5pt}
\end{figure*}

\vspace{-15pt}
\subsection{Measurment of reflection phase vortex}
\vspace{-10pt}
We have developed an efficient method to experimentally detect the ultrarobust reflection phase vortices, which outperforms the usual angle-resolved spectrometry by enabling the acquisition of reflection spectra at nearzero angles with ease. 
Figure~\ref{Fig-4}a depicts the diagram of our setup, wherein a point source is placed on the left of the GDZIM sample and microwave absorbers are tiled around the periphery. Using a network analyzer, we first probed the total electric field $E^\text{prob}_z(\mathbf{r}) = E^\text{in}_z(\mathbf{r}) + E^\text{r}_z(\mathbf{r})$ spreading in the region between the source and the sample. Using discrete Fourier transform (see Supplementary Note S6.A), we then obtained the plane-wave components, $\tilde{E}^\text{prob}_z(\mathbf{k})$, of the probed field distributed on the light cone of free space (Fig.~\ref{Fig-4}b). 
The plane-wave components with opposite signs of $k_x$ have clearly different origins: all the $k_x>0$ components represent the fields that are directly emitted from the source and incident upon the GDZIM slab, while all the $k_x<0$ components correspond to the fields reflected back from the sample. Hence, for any given $k_y$, the reflection coefficient can be extracted as $R(k_y,\omega) = \tilde{E}^\text{prob}_z(-|k_x|,k_y)/\tilde{E}^\text{prob}_z(|k_x|,k_y)$ with $|k_x|=\sqrt{(\omega/c)^2-k_y^2}$. By scanning across the frequency range $8-12$ GHz, we finally obtained the reflection spectra exhibited in Fig.~\ref{Fig-4}d for four samples with differing layer numbers 1 to 4. The measured spectra of all samples manifest, with sufficient accuracy, that a reflection phase vortex always arises at the projected Dirac point irrespective of the thickness of the GDZIMs. These agree well with the theoretical results presented in Fig.~\ref{Fig-4}c and thus demonstrate the ultrarobustness of the topological vortex stemming from the bulk topological phase transition.

\vspace{-15pt}
\subsection{Intrinsic generation of spatiotemporal vortex pulses}\vspace{-10pt}
With the vortical reflection phase, GDZIM slabs provide an intrinsic route towards robustly generating STVPs through reflecting a normally incident Gaussian pulse $E_z^\mathrm{in}(\mathbf{r},t)=A^\mathrm{in}(\mathbf{r},t)e^{i\omega_0(x/c-t)}$ 
whose central frequency is the Dirac frequency, $\omega_0$, of the GDZIM. 
The time slices in Fig.~\ref{Fig-5}a show the generation of the reflected STVP from a homogeneous GDZIM slab. 
Theoretically, the $(k_y,\omega)$-plane vortical reflection spectrum yields an approximate expression of the reflected pulse~\cite{Wang:21} (see Supplementary Note S5)
\begin{equation}
\begin{aligned}
     &E_z^\mathrm{r} (\mathbf{r},t) \propto d\left[\frac{c\kappa(x+ct)}{\Delta x^2} -i\frac{y}{\Delta y^2}\right]A^\mathrm{in}(\mathbf{r},-t)e^{-i\omega_0(x/c+t)},
\end{aligned}
\end{equation} 
which exposes the emergence of a spatiotemporal vortex along the centroid worldline of the reflected pulse, $\mathbf{r}_\mathrm{c}=-ct\,\hat{\mathbf{x}}$, in real spacetime.
Hence, either on a spatial cross-section ($t=\mathrm{const.}$) or on a spacetime cross-section (e.g., the plane of $x=\mathrm{const.}$ in Fig.~\ref{Fig-5}c) that traverses the worldline of the pulse, a 2D phase vortex can be observed. From the perspective of topology, the generation of the STVP is the ultimate consequence of the BBC in Eq.~\eqref{BBC}, thereby intrinsically mirroring the bulk critical Dirac topology of GDZIMs. From the perspective of angular momentum transport, the transverse OAM of the reflected STVP originates essentially from the transverse spin angular momentum of the eigenstates in GDZIMs.

Based on this approach, we have successfully generated and observed STVPs using the GDZIM slabs, with an experimental setup schematically shown in Fig.~\ref{Fig-5}b.  
We first utilized an arbitrary waveform generator(AWG) to emit a temporal Gaussian pulse with cylindrical wavefronts. After passing through a 3D-printing metamaterial lens (see Supplementary Note S6.C), the pulse is reshaped into a Gaussian-like wavepacket that propagates along the $x$-direction and then strikes upon the GDZIM sample. Using a real-time oscilloscope, we recorded the time series of the incident and reflected pulses along a scan line in the $y$-direction (the dashed line). 
To ensure the synchronization of signals measured at different points on the scan line, a reference signal, which is synchronous with the pulse-input channel, was generated from the AWG and used as the trigger signal for the oscilloscope. We have detected the STVPs produced by the PC samples of 1 to 4 layers, respectively. The experimental measurements, after removing the fast oscillatory dynamical phase $A(\mathbf{r},t)=E_z(\mathbf{r},t)e^{i\omega_0t}$, are compared with 
the simulated results of both effective media and metamaterials, as shown in Figs.~\ref{Fig-5}c-e. We can see that spatiotemporal phase vortices appear on the $(t,y)$ spacetime cross-section for all samples, and they exhibit a consistent geometric phase distribution, $\arg(A(\mathbf{r},t))$, around the vortex centers, though the strength of reflected pulses increases with the slab thickness. This demonstrates that the generation of the STVPs is totally insensitive to the variation in the thickness of the sample.

\vspace{-15pt}
\subsection{Conclusion and outlook}\vspace{-10pt}
By constructing topological metamaterials residing at the critical transition point of photonic Chern insulator phases, we have experimentally realized GDZIMs, a conceptual advancement over the conventional DZIMs, that incorporates nonreciprocal and tensor-valued constitutive indices of null determinants. The nonreciprocal nature of GDZIMs, marked by an unpaired spin-1/2 Dirac point at the center of the momentum space, gives rise to a striking spatiotemporal vortical reflection spectrum securely pinned at the surface projection of the bulk Dirac point. Remarkably, we established a new form of BBC, which unveils that the reflection phase vortex is essentially determined by the change of the bulk topological invariants across the phase transition and ultimately leads to an intrinsic and ultrarobust pathway to the generation of STVPs through the use of GDZIMs. The realization of GDZIMs not only opens up boundless promises for achieving highly functional devices, such as unidirectional ZRI waveguides~\cite{davoyan2013Optical} and scalable single-mode chiral emitting lasers~\cite{contractor2022Scalable}. It also builds a bridge that connects ZRI photonics~\cite{liberal2017Nearzero,kinsey2019Nearzeroindex}, topological photonics~\cite{ozawa2019Topological}, and singular optics~\cite{ni2021Multidimensional}, shedding new light on each field and prefiguring uncharted interdisciplinary research directions.

\bibliography{scibib.bib}

\vspace{20pt}
\textbf{Acknowledgements:}
The authors would like to thank Sir John Pendry, Profs. Kun Ding, Yun Lai, Guancong Ma, and Ying Wu for helpful discussions. The authors also thank Dr. Kam Man Shum for his support in the experiments.
This work is supported by the Research Grants Council of Hong Kong (16310422, AoE/P-502/20, and CityU 21207824) and by the National Natural Science Foundation of China (12174263).

\textbf{Author contributions:} R.-Y.Z., X.C., N.W., and C.T.C. conceived the idea. R.-Y.Z., X.C., W.L., and N.W. developed the theory. X.C. performed numerical simulations. X.C., R.-Y.Z., M.W., and D.W. designed and carried out the static experimental measurements. Y.-S.Z., X.C., G.-B.W., and J.C. designed and carried out the time-domain experiments. R.-Y.Z., X.C., Y.-S.Z., and C.T.C. wrote the manuscript. C.T.C., G.-B.W. and Z.-Q.Z. supervised the project.
All authors contributed to the discussions.

\textbf{Competing interests:}
The authors declare no competing interests.

\textbf{Data availability:}
Source data are provided with the paper. All other data that support the plots within the paper and other findings of the study are available from the corresponding author upon reasonable request.

\textbf{Code availability:}
The code used to evaluate the conclusions in the paper is available upon request.

\end{document}